# Less is more: sampling chemical space with active learning

*Justin S. Smith[1], Ben Nebgen[3], Nicholas Lubbers[3], Olexandr Isayev[2,\*], Adrian E. Roitberg[1,\*]*

[1]Department of Chemistry, University of Florida, Gainesville, FL 32611, USA

[2]UNC Eshelman School of Pharmacy, University of North Carolina at Chapel Hill, Chapel Hill, NC 27599, USA

[3]Los Alamos National Laboratory, Los Alamos, NM 87545, USA

\* Corresponding authors; email: OI (olexandr@olexandrisayev.com) and AER (roitberg@ufl.edu)

**ABSTRACT**

The development of accurate and transferable machine learning (ML) potentials for predicting molecular energetics is a challenging task. The process of data generation to train such ML potentials is a task neither well understood nor researched in detail. In this work, we present a fully automated approach for the generation of datasets with the intent of training universal ML potentials. It is based on the concept of active learning (AL) via Query by Committee (QBC), which uses the disagreement between an ensemble of ML potentials to infer the reliability of the ensemble's prediction. QBC allows the presented AL algorithm to automatically sample regions of chemical space where the ML potential fails to accurately predict the potential energy. AL improves the overall fitness of ANAKIN-ME (ANI) deep learning potentials in rigorous test cases by mitigating human biases in deciding what new training data to use. AL also reduces the training set size to a fraction of the data required when using naive random sampling techniques. To provide validation of our AL approach we develop the COMP6 benchmark (publicly available on GitHub), which contains a diverse set of organic molecules. Through the AL process, it is shown that the AL-based potentials perform as well as the ANI-1 potential on COMP6 with only 10% of the data, and vastly outperforms ANI-1 with 25% the amount of data. Finally, we show that our proposed AL technique develops a universal ANI potential (ANI-1x) that provides accurate energy and force predictions on the entire COMP6 benchmark. This universal ML potential achieves a level of accuracy on par with the best ML potentials for single molecule or materials, while remaining applicable to the general class of organic molecules comprised of the elements CHNO.

# I. INTRODUCTION

The development of accurate force fields[1–3] for the efficient simulation of large and small molecular systems has been a cornerstone of modern computational chemistry. The popularity of force fields is driven by low computational cost relative to more accurate and transferable quantum mechanical (QM) methods, such as density function theory[4] (DFT) or post-Hartree-Fock[5–7] methods. However, parametrizing *universal* force fields--applicable to any chemical system in any chemical environment--has remained an elusive goal due to the restrictive functional form and tedious atom typing of classical force fields. For this reason, a "zoo" of force fields has been developed over the last 30 years with applications in various regions of chemistry and physics, such as materials, proteins, carbohydrates, and small drug-like molecules.[8–11] Drawing a line between where these system-specific force fields work and where they fail is a challenging task.

In recent years, machine learning (ML) methods have been successfully applied in many areas of chemistry and physics research.[12–19] Specifically, ML approaches for the prediction of interatomic potential energy surfaces (referred to as ML potentials) have exhibited chemical accuracy compared to QM models at roughly the computational cost of classical force fields.[20–31] ML potentials promise to bridge the speed vs. accuracy gap between force fields and QM methods. Many recent studies rely on a philosophy of parametrization to one chemical system at a time[22,25], single component bulk systems[27,28] or many equilibrium structures, i.e. QM7 and QM9 datasets[32,33]. While parametrization to one system at a time can achieve high accuracy with relatively small amounts of QM, it has the downside that one must generate additional QM data and train a new ML model for every new chemical system. Using this approach in any study requires extra parametrization time due to the non-universality of the potentials. Additionally, parametrization to only equilibrium geometries does not attempt to describe the range of conformations visited in atomistic simulation. For these reasons, single system and equilibrium dataset ML potentials do not aim to build an extensible and transferable (universal) ML potential.

Our work on the ANAKIN-ME (ANI) method for developing the ANI-1 potential[34] is one example of a universal ML atomistic potential for organic molecules. The methodology is built upon the concept of an atomic environment descriptor first developed by Behler and Parrinello[35] and refined to perform significantly better on large and diverse datasets of organic molecules. A key aspect of the ANI methodology was the focus on dataset diversity, which promotes the learning of low level interactions (by utilizing localized descriptors) for better transferability. For training the ANI-1 model, we calculated over 22 million structural conformations from 57,000 distinct small organic molecules using DFT.[36] The ANI-1 dataset was built through an exhaustive sampling of molecules containing between one and eight C, N, and O atoms from the GDB-11 database, with H atoms added to saturate the configurations. The ANI-1 dataset is built on a philosophy of dataset construction that samples small molecule *conformational and configurational space* at the same time. The ANI-1 potential was shown to be chemically accurate for systems of 50 atoms and more, demonstrating extensibility and transferability to much larger molecules than those in the training set. This phenomenon, whereby an ML model is trained on small systems (which could be thought of as fragments of large systems), then demonstrated to be extensible to large systems has also been confirmed in other recent studies.[29,37,38] Other recent work had success in developing universal ML property predictors for organic based chemical systems away from their local minima.[29]

When it comes to developing or optimizing ML model training datasets, human intuition currently drives the experiment design. The resulting datasets tend to be clustered, sparse, and incomplete; recent work finds that people tend to favor inclusion of "successful" experiments and tend to forget "failed" experiments.[39] The comprehensive incorporation of all data is the strength of ML approaches to artificial intelligence (AI). With sufficient data, an AI-driven machine can more effectively choose the next step in experiments or simulations than humans, speeding up the optimization of a given dataset, while also reducing the overall amount of data required. As robotics transforms chemical synthesis[40], manufacturing, and transportation, constituting a modern industrial revolution,[41,42] achieving the analogous revolution in computational methods will require AI and in particular the emulation of scientific intuition, reasoning and decision making. Such an ambitious program will not be accomplished all at once and will instead require incremental progress as AI algorithms are developed.

In this work we present a fully automated approach of dataset generation for training universal ML potentials. It is based on the concept of active learning (AL), which has been successfully applied to develop single system ML potentials[37,43–46] and in other areas[47,48] of chemical sciences. We develop a two-component technique for training universal ML potentials. The first component is a dataset reduction algorithm for eliminating redundancy in an existing training set. The second is an active learning algorithm based on the query by committee[49] (QBC) approach for selecting new training data. For a complete and rigorous validation of universal potentials, we also develop the COmprehensive Machine-learning Potential (COMP6) benchmark suite for organic and bio-molecules. The COMP6 benchmark samples the chemical space (for molecules containing C, H, N, and O) of molecules larger than those included in the training set, as well as non-covalent interactions via the S66x8 benchmark[50]. The COMP6 benchmark is publicly available on GitHub [https://github.com/isayev/COMP6]. Using the active learning scheme, a potential can be trained to the accuracy of ANI-1 using 90% less data, even while sampling from smaller molecules. After further exploration of chemical space, our potential (dubbed ANI-1x) strongly out-performs ANI-1, while being trained on a dataset that is only 25% of the size.

## II. METHODS

In the context of this work, the goal of active learning is to infer an accurate predictor from labeled training data. These labeled data are input-output pairs ($X$, $y$), where the output $y$ represents the correct answer to a question associated with the input $X$. In the problem of ML potential training, the label $y$ may be the "yes" / "no" answer to whether the potential correctly describes a molecule X. As part of the active learning process, this question may be answered empirically for a given substance. The Query by Committee (QBC) approach uses the disagreement between models trained to similar data to experimentally infer the correctness of an ensemble's prediction. This is by the following reasoning: if an ensemble of predictors has a high variance, then some models in the ensemble must have a relatively high error from the ground truth. Therefore, selection of compounds that have a high variance of ensemble predictions in search of new molecules and conformation can be employed to sample high error regions chemical space automatically, minimizing the need for redundant QM calculations. Several studies provided empirical evidence that this method of sampling indeed improves the overall fitness of ML potentials for single systems.[37,51] In this work, we apply this concept in a massive search of chemical space to develop a superior training set for universal ML ANI[34] potentials. These ANI potentials are applicable to organic molecules containing C, H, N, and O. With minimal modification, the same approach could be used for other areas of chemical sciences, e.g. materials.

## A. Sample selection via Query by Committee

We show how, in a rigorous statistical way, one can obtain *a priori* information about what new samples should be included in subsequent generations of an ML potential training set. The *a priori* information is obtained by the QBC[49] algorithm. QBC measures the disagreement between students (models) of a committee (ensemble), then the algorithm selects new examples where the students disagree by a preset inclusion criterion. Finally, new reference data for selected examples are obtained and included in the next committee training iteration. As a test of agreement, we choose to include new data point $i$ only for test cases which generate a value $\rho_i$ greater than an inclusion criterion $\hat{\rho}$, where $\rho_i$ is defined as,

$$\rho_i = \frac{\sigma_i}{\sqrt{N_i}} \qquad 1.$$

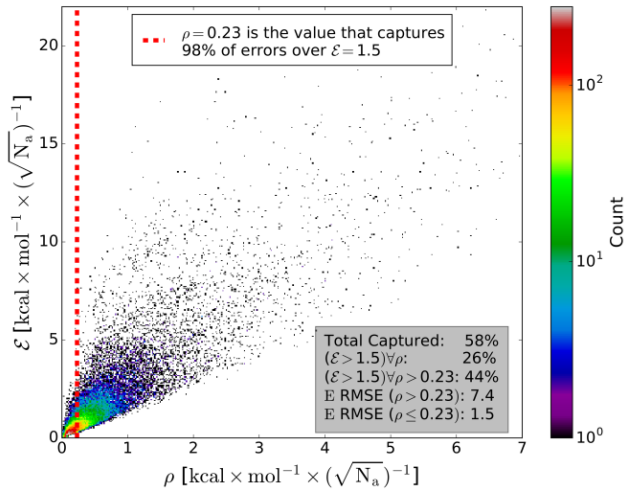

**Figure 1**: Example of choosing a value $\hat{\rho}$ which captures 98% of all errors ($\varepsilon$) over 1.5 kcal/mol on the GDB07to09 benchmark set using the initial (before using active learning) ANI model ensemble. The value which accomplished this is found to be $\hat{\rho} = 0.23$. This value of $\hat{\rho}$ used in query by committee results in the selection of 58% of all test data. Initially 26% of all $\varepsilon$ are greater than 1.5. 44% of $\varepsilon$ corresponding to $\rho > \hat{\rho}$ are greater than 1.5. Splitting the dataset along $\rho = \hat{\rho}$ results in a total energy RMSE of the ANI ensemble prediction vs. reference DFT of 7.4 kcal/mol for all values $\rho > \hat{\rho}$ and 1.5 kcal/mol for all values $\rho \leq \hat{\rho}$.

In equation 1, $\sigma_i$ is the standard deviation of predictions from an ensemble (see Section IIE for details) of ANI potentials and $N_i$ is the number of atoms in the given test system. The square root is applied to $N_i$ since the potentials are atomistic, and the total energy error is assumed to be a random distribution, centered around zero, per atom. That is, cancellation of error on a per atom basis can lead to artificially low per atom errors (and standard deviations in this case) on larger molecules when a square root is not applied. This is necessary when using a single value of $\hat{\rho}$ to test across molecules with varying numbers of atoms as is done in this work.

Figure 1 provides an example of how the inclusion criterion $\hat{\rho}$ is determined. In this 2-dimensional density plot, $\varepsilon_i = |MAX(\{E_T^{ANI}\}_i^{ens} - E_{T,i}^{REF})|/\sqrt{N_i}$ where $N_i$ is the number of atoms in the i$^{th}$ molecule. Therefore, $\varepsilon_i$ is the largest per atom prediction error of any model in the ensemble of ANI models for test molecule $i$. The test data used in this example is the GDB07to09 test set, which is described in Section IIC. The ANI model used to determine $\hat{\rho}$ in this example is the ANI model which initialized the AL process (Section IIB). The value $\hat{\rho}$ is determined from the choice of what value of $\varepsilon$ is considered too large, and what percentage of epsilon over that should be considered as fail cases. Therefore, $\hat{\rho} = 0.23$ was empirically selected as it is the value which allows selection of 98% of all $\varepsilon_i > 1.5$ kcal/mol.

The example from Figure 1 determines $\hat{\rho} = 0.23$ kcal/mol leads to the selection of 58% of all test data as molecules that fail the agreement test. As evidence that the choice of $\varepsilon_i$ allows for the statistical determination of poorly fit data, it is shown that before selecting any data (i.e. for all $\rho_i$), 26% of the complete test set $\varepsilon_i$ are greater than 1.5. However, this is 44% when considering all $\varepsilon_i > 1.5$ kcal/mol which correspond to $\rho_i > \hat{\rho}$. This shows that the determined $\hat{\rho}$ leads to a selection of data with a greater number of $\varepsilon_i > 1.5$ kcal/mol within its population. As further validation of the approach, the application of the concept is shown to choose "bad" data by calculating the RMSE of the potential energy ($E$) for the mean prediction of the ensemble of ANI models vs. reference DFT calculations. For all $i$ molecular structures corresponding to $\rho_i > \hat{\rho}$, the $E$ RMSE is 7.4 kcal/mol. On the other hand, for all $i$ molecular structures corresponding to $\rho_i \leq \hat{\rho}$ the $E$ RMSE is 1.5 kcal/mol. Therefore, in a statistical way, the method chooses new data which is significantly higher in error compared to GDB07to09, which is randomly generated data.

With enough processing time on HPC resources, the rate-limiting step of a QBC data selection cycle using ANI potentials is the training of a new ensemble of ANI models. Complete training of a single network takes 40 minutes per one million data points on a single NVIDIA Tesla V100 GPU. To reduce the number of models trained, QBC is used in batches, searching configurational and conformational (chemical) space for tens of thousands of new reference data points that fail the agreement test. Finally, labels (reference potential energies, $E^{REF}$) are computed for all molecules in the selected batch. This process may lead to some redundant data. However, the alternative, retraining a new model ensemble after the addition of every new data point, will be impractically slow.

**B. Automatic chemical space sampling via active learning**

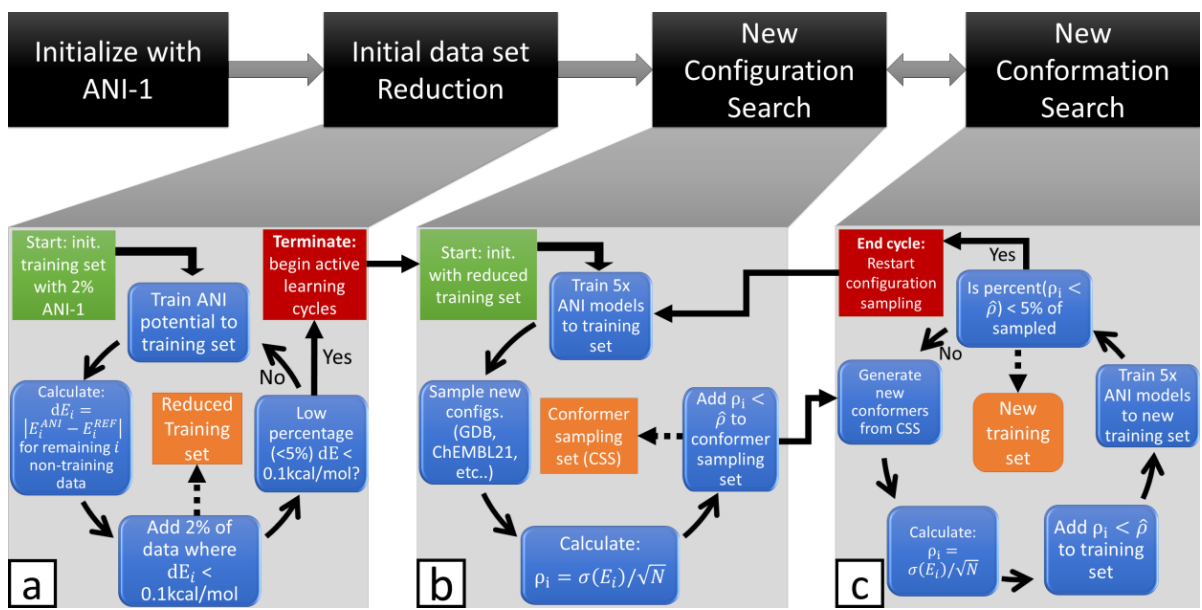

**Figure 2.** Fully automated AL workflow for data generation. The algorithm contains 3 steps: a) an existing dataset reduction, b) a configurational search, and c) a conformational search.

Figure 2 shows the overall workflow of the iterative AL algorithm. The algorithm is initialized from an existing random sampling generated dataset, which may contain some amount of redundant data. This initial dataset (ANI-1 in this work) is then reduced through an iterative approach with the goal of minimizing the overall dataset size while not impacting predictive performance. The reduction algorithm is provided in detail in Figure 2a. Figure 2a is initialized with a random subsampled 2% of the original ANI-1 dataset. Then, iteratively, the remaining data are tested, and 2% subsets of the fail cases are added to the training set. Here, a fail case is defined as $|E_{ANI} - E_{DFT}|/\sqrt{N} > 0.04\ kcal/mol$, where N is the number of atoms in the molecule. The algorithm is terminated when less than 5% of the data not yet added to the training set are considered as fail cases. The remaining < 5% high error data are added to the final dataset. Hyper-parameters for the reduction algorithm can be tuned to further reduce redundancies in the data, at the cost of more cycles, and therefore, longer run time. The final reduced dataset is used to bootstrap the remaining cycles of the active learning algorithm. If a dataset such as ANI-1 is not available, this step can be replaced with the generation of a small amount randomly sampled data across many small, one to five C, N, O atoms, molecules. However, this will lead to more active learning cycles before achieving the desired result.

With the reduced dataset, the configurational search (Figure 2b) is initialized. The configurational search is carried out by randomly sampling an external database of small molecules (e.g. GDB-11[52,53], ChEMBL[54–56], algorithmically generated dipeptides using RDKit [www.rdkit.org], automatically generated dimers), embedding the molecule in 3D space with RDKit, then optimizing initial structure with the UFF[57] force field. See supplemental information Section S1.2.3 for details on dimer generation. Next, ANI energies are computed using an ensemble of five

ANI models trained to the current AL dataset (see Section IID for details on ensemble prediction and training). Finally, $\rho_i = \sigma_i/\sqrt{N}$ is computed. Here, $\sigma_i$ is the standard deviation of the ensemble's energy predictions for molecule i and $N$ is the number of atoms in the molecule. The test of whether to include the molecule corresponding to a given $\rho_i$ is $\rho_i > \hat{\rho}$. The selection of $\hat{\rho}$ is explained in Section IIA. All molecules that fail this test are included in the new conformer sampling set. Any molecules added to the conformer sampling set are geometry optimized with the correct reference QM level of theory using tight SCF and optimization convergence criteria.

With the configurational search complete, a conformational search cycle (Figure 2c) is initialized, whereby the conformer sampling set (a set of equilibrium molecules generated in the configuration sampling step) is used to generate a set of new non-equilibrium molecules ($\hat{X}$). The conformers in $\hat{X}$ are generated via one of three techniques, which are designed to sample various regions of chemical space. These sampling techniques are listed below.

- Diverse normal mode sampling (DNMS). A version of normal mode sampling (NMS) as presented in our previous work[36], but with diversity selection used to reduce redundant data and a bias towards near equilibrium structures. A detailed description of DNMS is provided in supplemental information (SI) section S1.2.1.
- K random trajectory sampling (RTS). We run short (4 ps) molecular dynamics simulations, with an ensemble of ANI networks, starting with random velocities equal to 300K and heated slowly to 1000K over the simulation time. During the simulation, each step QBC is used to check whether the current structure fails the agreement test. Once a structure is reached that fails the test the simulation is terminated, and new QM data is generated for inclusion in the training set. This is repeated to generate multiple new samples. A detailed description of RTS is provided in SI section S1.2.2.
- Molecular dynamics generated dimer sampling. Dimers are generated by randomly placing and orienting molecules from the conformer sampling set into a box with periodic boundary conditions. A molecular dynamics simulation for 5ps is then carried out on the box. Every 50 steps the box is fragmented into only dimer pairs within the desired cutoff radius. Each new dimer pair is tested using the QBC approach, failed tests are kept as new data, and QM properties are generated for inclusion in the training set. A detailed description of the dimer sampling approach used here is provided in SI section S1.2.3.

After new data is selected, labels are computed and included in the training set, a new ensemble of ANI potentials is trained. The conformational search cycles are repeated until the model stops improving within the COMP6 benchmarks (see details in Section IIC). Finally, the entire cycle is restarted from the configurational sampling step. This process is carried out to produce a total of 37 cycles including many configurational and conformational searching cycles. Throughout this work we will refer to various intermediate active learned ANI models as AL1 through AL6. The AL6 potential is the final potential reached in this work and is referred to as the ANI-1x potential, which is provided for free in a python package integrated with the atomic simulation environment (ASE) package[58] [https://github.com/isayev/ASE_ANI]. The first row in Table 1 provides information about the final dataset from this work, labeled as ANI-1x. Notably, the size of the ANI-1x dataset, at 5.5 million structures, is 25% the size of the dataset used in training the original ANI-1 potential (22M).

**C. Development of the COMP6 benchmark suite**

**Table 1:** Description of the final active learning generated training dataset (ANI-1x) and all six COMP6 benchmark datasets. Mean relative energy range is the average range of relative energies for each set of conformers. Energy prediction range is the real prediction range in the benchmark; this is the range which the ANI model predicts energies in. [energy units: kcal/mol]

| Purpose | Dataset | Molecule Source | Configurations (Conformations) | Atoms/Molecule mean (std. dev.) | Mean Relative Energy Range | Energy Prediction Range |
|---|---|---|---|---|---|---|
| Training | ANI-1x | ANI-1 + AL | 63,865 (5,496,771) | 15 (5) | 97.6 | 6,400 |
| Testing | S66x8 | S66x8 | 66 (528) | 20 (7) | 6.00 | 2,800 |
| | ANI-MD | PDB | 14 (1,791) | 75 (72) | 35.0 | 31,000 |
| | GDB7to9 | GDB-11 | 1,500 (36,000) | 17 (3) | 78.0 | 1,900 |
| | GDB10to13 | GDB-13 | 2,996 (47,670) | 25 (4) | 214.0 | 2,300 |
| | Tripeptides | RDKit | 248 (1,984) | 53 (7) | 102.0 | 4,200 |
| | DrugBank | DrugBank | 837 (13,379) | 44 (20) | 167.0 | 14,000 |

To validate that the active learning process generates an ANI potential which outperforms the original ANI-1 potential, and that each cycle's resulting AL ANI potentials consistently outperforms previous versions of AL ANI potentials, we develop the comprehensive machine learned potential (COMP6) benchmark. COMP6 is a benchmark suite composed five rigorous benchmarks that cover broad regions of organic and bio-chemical space (for molecules containing C, N, O, and H atoms) and a sixth built from the existing S66x8[50] noncovalent interaction benchmark. The five new benchmark sets are referred to as GDB7to9, GDB10to13, Tripeptides, DrugBank, and ANI-MD. See Table 1 for a detailed description. The benchmarks range from a mean molecule size of 17 atoms to 75 atoms, with the largest molecule being 312 atoms. Below is a description of the methods used to develop each benchmark. Energies and forces for all non-equilibrium molecular conformations presented have been calculated using the ωB97x[59] density functional with the 6-31G(d) basis set[60] as implemented in the Gaussian 09[61] electronic structure software. Hirshfeld charges and molecular dipoles are also included in the benchmark. An analysis of these properties will be carried out in future work.

- S66x8 Benchmark. This dataset is built from the original S66x8[50] benchmark for comparing accuracy between different methods in describing noncovalent interactions common in biological molecules. S66x8 is developed from 66 dimeric systems involving hydrogen bonding, pi-pi stacking, London interactions, and mixed influence interactions. While the keen reader might question the use of this benchmark without dispersion corrections, since dispersion corrections such as the D3[62] correction by Grimme et al. are *a posteriori* additions to the produced energy, then a comparison without the correction is equivalent to a comparison with the same dispersion corrections applied to both models.
- ANI Molecular Dynamics (ANI-MD) Benchmark. Forces from the ANI-1x potential are applied to run 1ns of vacuum molecular dynamics with a 0.25fs time step at 300K using the Langevin thermostat on 14 well-known drug molecules and two small proteins. System sizes range from 20 to 312 atoms. A random subsample of 128 frames from each 1ns trajectory is selected, and reference DFT single point calculations are performed to obtain QM energies and forces.

- GDB7to9 Benchmark. The GDB-11 subsets containing 7 to 9 heavy atoms (C, N, and O) are subsampled and randomly embedded in 3D space using RDKit [www.rdkit.org]. A total of 1500 molecule SMILES [opensmiles.org] strings are selected: 500 per 7, 8, and 9 heavy-atom set. The resulting structures are optimized with tight convergence criteria, and normal modes/force constants are computed using the reference DFT model. Finally, diverse normal mode sampling (DNMS) is carried out to generate non-equilibrium conformations.
- GDB10to13 Benchmark. Subsamples of 500 SMILES strings each from the 10 and 11 heavy-atom subsets of GDB-11[52,53] and 1000 SMILES strings from the 12 and 13 heavy-atom subsets of the GDB-13[63] database are randomly selected. DNMS is utilized to generate random non-equilibrium conformations.
- Tripeptide Benchmark. 248 random tripeptides containing H, C, N, and O are generated using FASTA strings and randomly embedded in 3D space using RDKit. As with GDB7to9, the molecules are optimized, and normal modes are computed. DNMS is utilized to generate random non-equilibrium conformations.
- DrugBank Benchmark. This benchmark is developed through a subsampling of the DrugBank[64] database of real drug molecules. 837 SMILES strings containing C, N, and O are randomly selected. Like the GDB7to9 benchmark, the molecules are embedded in 3D space, structurally optimized, and normal modes are computed. DNMS is utilized to generate random non-equilibrium conformations.

**D. Error metrics for validation on the COMP6 benchmark suite**

This work uses three error metrics for comparing different versions of ANI potentials: potential energy ($E$), conformer energy difference ($\Delta E$), and atomic force component errors ($F$).

- Potential energy ($E$) error is a comparison of $E_i^{M1}$, potential energies produced by model M1 for molecule i, to $E_i^{M2}$, the potential energies produced by model M2 for molecule i.
- The conformer energy difference ($\Delta E$) error is calculated per set of conformers. In the benchmark dataset K sets of conformers are supplied, one per molecular configuration. For a given set of conformers k, the conformer energy difference between conformers $i$ and $j$ where for a given model M is obtained by computing $\Delta E_{ij}^{M,k} = E_i^{M,k} - E_j^{M,k}$. Finally, error is calculated between $\Delta E_{ij}^{M1,k}$ and $\Delta E_{ij}^{M2,k}$ for all k, $i$, and $j > i + 1$ for models M1 and M2.
- The atomic force ($F$) error metric is a comparison between the individual components (x, y, z) of each atom's force vector for all conformations included in the given benchmark.

Comparisons are given in mean absolute error (MAE), and root mean squared error (RMSE) throughout this article. The comparison of MAE along with RMSE can give information about outliers in a model's predictions. For example, two models can have the same MAE for a prediction on a given benchmark while the RMSE can be much higher for one than the another. For this reason, it is good practice to provide both MAE and RMSE when comparing two methods on some benchmark.

**E. Property prediction with an ensemble of ANI models**

For energy and force predictions we use the mean prediction of an ensemble of ANI potentials. The concept of using an ensemble mean for ML model prediction is common practice in the ML community. Recently, it has been adopted in the area of ML molecular property prediction.[31,37,65] All potentials used to generate results in this work utilize the mean prediction for an ensemble of $L = 5$ ANI potentials trained to a 5-fold cross validation split of the training dataset. The potential energy ($E$) is represented by,

$$E = \frac{1}{L} \sum_{i=1}^{L} E_i$$

where $E_i$ is the potential energy prediction from each of an ensemble's $L$ ANI models. Since the models are independent, atomic forces for the ensemble can be derived as the component wise mean of the forces from the $L$ individual ANI models. The use of an ensemble as described above decreases ANI vs. DFT $E$ RMSE by 0.67 kcal/mol, $\Delta E$ RMSE by 0.68 kcal/mol, and $F$ RMSE by 2.1 kcal/mol $\times$ Å$^{-1}$ over the entire COMP6 benchmark, an error reduction of 17%, 19% and 28%, respectively.

## III. RESULTS AND DISCUSSIONS

The supplemental information (SI) provided with this work contains various tables detailing the results obtained on the COMP6 benchmark by the ANI potentials discussed in this work. SI tables S1 through S7 provide an analysis of the $\Delta E$, $E$, and $F$ errors obtained for six subsequent active learned ANI potentials, AL1 through AL6, and the original ANI-1 potential. Note that the publicly released ANI-1x potential is the AL6 ANI potential. SI Tables S8 through S10 supply an analysis of the individual ANI-MD trajectory results for the ANI-1x potential. Table S9 provides per atom energy errors for the ANI-1x potential vs. DFT and shows that the mean energy prediction RMSE per atom for all trajectories is 0.05 kcal/mol per atom. This level of accuracy is on par with single molecule or bulk metal ML potentials as described in recent work by J. Behler.[25] SI Table S11 provides details on the ANI models introduced in this work. Finally, SI Tables S12 through S17 give errors for COMP6 considering conformers within select energy ranges for the ANI-1x potential. These tables show much lower errors for conformations which are thermally accessible to room temperature molecular dynamics simulations. As shown in Table S17, thermally accessible conformations (within 50kcal/mol) have a $E$ MAE/RMSE of 0.064/0.105 kcal/mol per atom and $\Delta E$ MAE/RMSE of 0.049/0.070 kcal/mol per atom over the complete COMP6 benchmark.

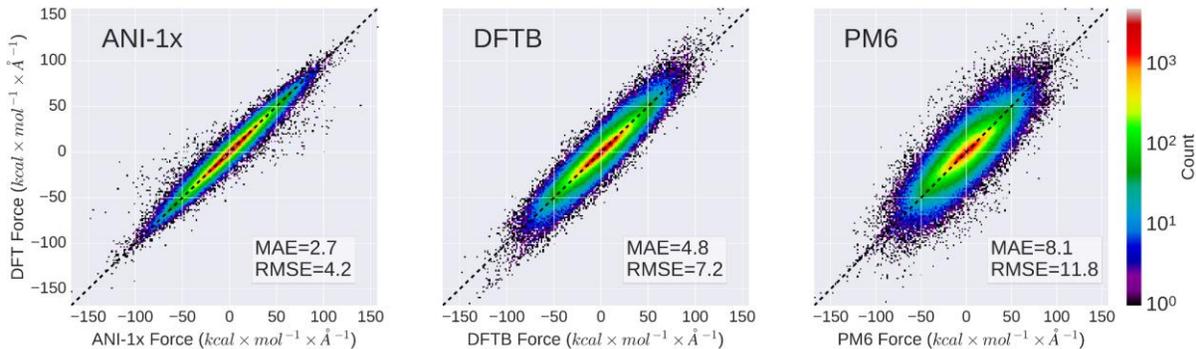

**Figure 3.** Force correlation plots comparing ANI-1x, DFTB (3ob-3-1 parameter set for bio-molecules), and PM6 to DFT reference calculations are provided from left to right, respectively, for the complete ANI-MD benchmark. Molecules in the ANI-MD benchmark are composed of a mean of 75 atoms with the largest being Trp-cage (1L2Y), a 20-residue (312-atom) protein. DFTB and PM6 are provided as a baseline of comparison. Mean absolute errors (MAE) and root mean squared errors (RMSE) are provided in the bottom right of each figure. The color bar scale is the same for all figures allowing a proper density comparison.

Figure 3 provides evidence of the ANI-1x force prediction capabilities. Also, most tables in the supplemental information further establish the accuracy of ANI potential force prediction on the COMP6 benchmark suite. By construction, ANI potentials provide analytic and energy-conservative forces, a requirement for molecular dynamics simulations. It is noteworthy that force training, which can be computationally expensive, **is not required** to achieve these force prediction results. The forces compared in the DFT correlation density plots in Figure 3 are from all trajectories combined in the COMP6 ANI-MD benchmark. We compare the same figures for ANI-1x (left), DFTB (center), and PM6 (right). DFTB and PM6 are included as a baseline for the comparison. The ANI-MD benchmark is a rigorous test case for any ML potential's force prediction because the molecules supplied in the dataset range from 20 to 312 atoms, with an average size of 75 atoms. A breakdown of the errors for each trajectory in the ANI-MD benchmark is supplied in the supplemental information, Tables S8 through S10.

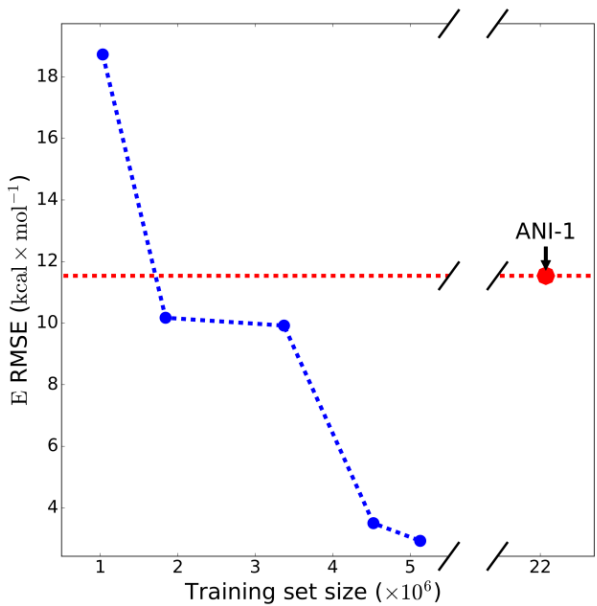

**Figure 4.** Comparison of potential energy (E) RMSE obtained on the entire COMP6 benchmark vs. training set size (total molecular conformation included in the training set). The x-axis represents the progression of the active learning process. Plot points are obtained by ANI potentials (blue) trained to various versions of the active learned dataset and an ANI potential (red) trained on the original ANI-1 dataset.

The closest comparison in literature can be found in recent work on a system specific ML potential for an alanine tripeptide where a force RMSE of 3.4 kcal/mol × Å$^{-1}$ was achieved with test data from a 350K MD trajectory.[37] The force error from this work was obtained by training directly to energies and analytic forces from fragments of the molecule being tested. In the case of the ANI-1x potential, which was used to predict the forces for the creation of the ANI-MD benchmark, a MAE/RMSE of 2.7/4.2 kcal/mol × Å$^{-1}$ is obtained vs. *a posteriori* DFT calculations on 128 random frames from each of the 14 molecule's 1ns molecular dynamics trajectories. More impressive, $F$ MAE/RMSE for the neutralized 20-residue Trp-cage (1L2Y) and 10-residue Chignolin (1UAO) proteins are 3.1/4.6 kcal/mol × Å$^{-1}$ and 3.3/4.7 kcal/mol × Å$^{-1}$, respectively. ANI-1x also exhibits a force MAE/RMSE of 2.3/3.3 kcal/mol × Å$^{-1}$ within the energy range of 50 kcal/mol on the tripeptide benchmark (non-equilibrium conformations from 248 randomly generated tripeptides) from COMP6 (see supplemental information Table S14), which is roughly the accessible energy range of 350K molecular dynamics simulations. Finally, considering the ANI-1x potential was utilized to generate 1 ns of 300K molecular dynamics simulations, from which the ANI-MD benchmark geometries were sampled, speaks to the applicability of the forces in molecular dynamics for general molecular systems. All the previously mentioned results from the ANI-1x potential were obtained **without** the need of direct force training.

Figure 4 provides a plot of *E* RMSE achieved on COMP6 vs. dataset size for various active learned datasets and the original ANI-1 dataset. With only 2 million data points, the active learned ANI potentials already outperform the original ANI-1 potential across the entire COMP6 benchmark. Once the active learned ANI potential reaches 5.5 million data points it five times outperforms ANI-1 and is approaching chemical accuracy from the reference DFT calculations. In the new COMP6 benchmark, diversity selection in the normal mode sampling helps ensure a more uniform sampling of energy states within the energy range being fit to and tested within. Therefore, general errors on COMP6 vs. the ANI-1 potential's original results are expected to be much higher on this complex benchmark than the results published on the less rigorous test sets from the original ANI-1 work. Table 1 provides the average energy ranges for each benchmark in COMP6 and the final training set (ANI-1x), as well as the energy prediction (atomization energy) range.

Most benchmarks in COMP6 (all but the ANI-MD benchmark) were used during the active learning process to validate the improvement in accuracy and universality of new active learned ANI models. Figure 5 provides the learning curves for six intermediate active learned ANI potentials on each benchmark in COMP6. Supplemental information Table S11 provides information of the chemical space sampled in each of these datasets. The horizontal dashed lines in Figure 5 represent the original ANI-1 ensemble predictions on each of the benchmarks for the property corresponding to its color. AL1 is the ANI potential used to initialize the active learning process. It was trained to a reduced (Figure 2a) version of the one through six heavy atom subsets of the ANI-1 dataset. AL2 through AL6 are successive versions of the active learned ANI potentials. More details for each active learning cycle shown in Figure 5 is provided in SI Table S11. During the active learning process, small molecules (one to six C, N, and O atoms) were initially sampled, with the size of the molecules sampled gradually increased as the active learning process continued. AL3 is where the AL models begin to statistically match or outperform the original ANI-1 model in most metrics. It is notable that AL3 accomplished this feat while only having sampled 1.8 million conformations from molecules with up to 7-heavy atoms from GDB-11. This shows that the active learning techniques employed in this work sample chemical space far better than random sampling techniques. Especially considering the ANI-1 dataset includes 22 million conformations from larger, up to 8-heavy atom, molecules.

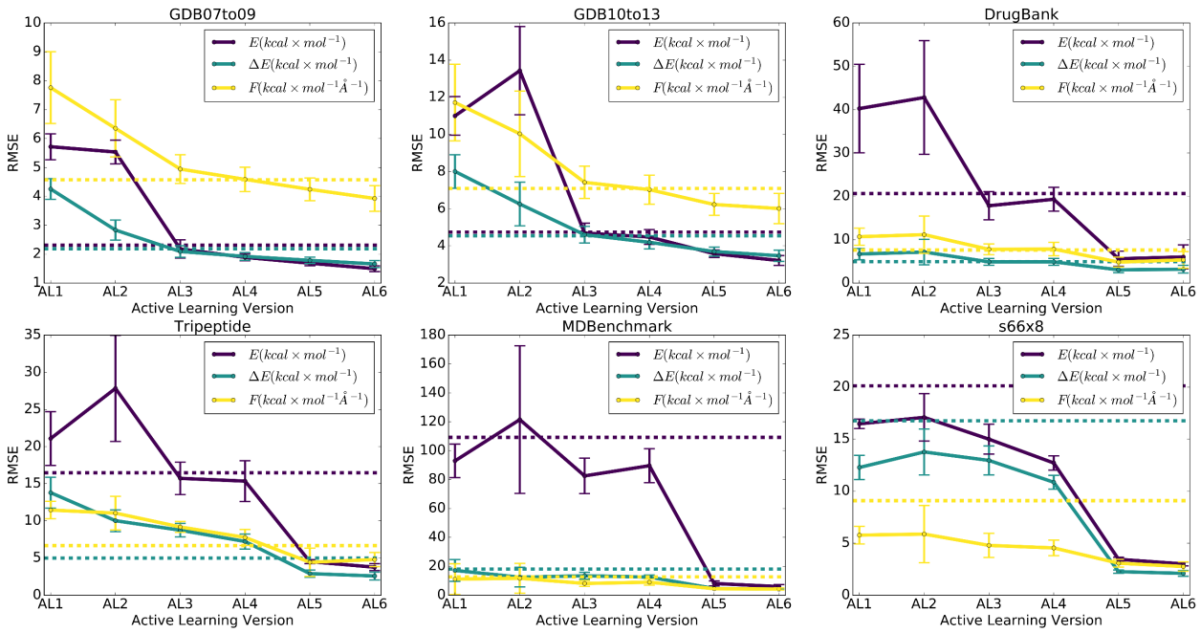

**Figure 5.** Individual COMP6 benchmark learning curves for successive versions of the active learned potentials. RMSE is provided for three properties: potential energy ($E$), conformer energy differences ($\Delta E$), and force components ($F$). The error bars on the solid lines represent one standard deviation of each of the five ANI models in the ensemble used to make the mean prediction. The horizontal lines represent the mean prediction of ANI-1.

Eventually, between the AL4 and AL5 steps, amino acids, generated dipeptides, generated small molecule dimers and small ChEMBL molecules were added to the sampling set. This is apparent from the large drop in error between AL4 and AL5 for the DrugBank, Tripeptides, and S66x8 benchmarks. Active learning sampling was also driven into the GDB-11's 9-heavy atom subset for sampling during the production of AL6. Supplemental information Tables S2 through S7 provide the full tables used to make Figure 5 along with Table S1 which describes all benchmarks at once. The latest ANI potential, ANI-1x (shown as AL6), achieves remarkable property prediction on the complete benchmark with errors (MAE/RMSE) of 1.9/3.4 kcal/mol ($E$), 1.8/3.0 kcal/mol ($\Delta E$), and 3.1/5.3 kcal/mol × Å$^{-1}$ (F) within the full energy range of the benchmark.

In general, as each ANI potential's fitness improves in Figure 5, the standard deviation (shown as vertical error bars) of each property prediction for a given ensemble decreases as well. This is a sign that each model in the ensemble is obtaining enough chemical interaction information through active learning that the models begin agreeing on their predictions for these larger systems. By the final iteration of the active learning cycles, an active learned dataset of 5.5M data points is used in training the ANI-1x potential. The ANI-1x potential outperforms the ANI-1 potential on all properties across all benchmarks. Further, the ANI-1x dataset is 25% the size of the original ANI-1 dataset, which contains a total of 22M data points.

There has been recent discussion in literature (Herr et al.[66]) questioning the validity of using data generated from a single sampling technique to successfully extrapolate to out-of-sample data. We believe such critique is well placed and has particular impact when defining system-specific potentials or other models of limited scope. In the present work, we combined several sampling techniques, attempting to cover all relevant regions of conformational and configurational (chemical) space. We test model performance on a separate and very diverse set of systems, showing not only accuracy, but extensibility to molecules and conformations much larger than the training set. Accuracies on benchmarks generated with different sampling techniques are comparably accurate: On the ANI-MD benchmark (mean energy range of 35 kcal/mol), we achieve force MAE of 2.49 kcal/mol × Å$^{-1}$ (SI Table S10). On the overall COMP6 benchmark restricted to the energy range of 50 kcal/mol, we achieve force MAE of 2.48 kcal/mol × Å$^{-1}$ (SI Table S12). The fact that MD-sampled test points show quite similar error to the mostly DNM-sampled benchmark is evidence that the active learning procedure with hybrid sampling methods produces a model that is robust.

**IV. CONCLUSIONS**

In pursuit of automated dataset generation for the development of universal machine learned potentials, we introduce automatic active learning techniques for sampling sparsely explored regions of chemical space. The algorithm begins with the reduction of an existing dataset to remove redundant data without loss of accuracy. New conformations of molecules are generated through normal mode sampling, molecular dynamics sampling, and random dimer sampling. Periodically the algorithm samples new molecular configurations from a variety of sources to diversify its exploration of chemical space. The result is a new potential (ANI-1x) developed though successive generations of the active learning process. The ANI-1x potential is packaged in a user-friendly Python library, which is publicly available on GitHub [https://github.com/isayev/ASE_ANI]. We also introduce the COMP6 benchmark for monitoring the progress of active learning cycles and for comparison to future universal potentials. The ANI-1x potential achieves errors (MAE/RMSE) of 1.6/3.0 kcal/mol ($E$), 1.4/2.3 kcal/mol ($\Delta E$), and 2.7/4.5 kcal/mol × Å$^{-1}$ (F) when testing on points within 100 kcal/mol of the energy minima for the complete COMP6 benchmark.

The COMP6 benchmark suite consists of six diverse benchmark test sets. The COMP6 benchmark suite is made publicly available for comparing future ML potentials [https://github.com/isayev/COMP6]. As provided, properties are calculated using the ωB97x density functional with the 6-31G(d) basis set, however, it could be recomputed using the desired quantum level of theory. For complete transparency, we provide the exact error metrics used to measure accuracy on the COMP6 benchmark suite. It is our hope that the COMP6 benchmark will provide the universal ML potential development community with a rigorous benchmark for comparison of ML potential methods on organic molecules in the extrapolative regime. The COMP6 benchmark suite constitutes a first benchmark of its kind for the comparison of universal ML potentials in this rapidly changing and ever-growing field.

The ANI-1x potential was trained to less than 100 conformations per molecular configuration in its training set, compared to 400 for the ANI-1 dataset. The accuracy of the ANI-1x potential is on par with the best single molecule or material ML potentials, while most single molecule parametrized ML potentials require many hundreds to thousands of conformations to parametrize a single system. This further validates the configurational and conformational big data sampling philosophy introduced in the original ANI-1 work. Since the mean molecule size in the ANI-1x active learning training set is only 15 total atoms (8 heavy atoms), the generation of more accurate post-Hartree-Fock datasets is now feasibles.

The high-level of universal accuracy achieved by the ANI-1x potential can be attributed to the capacity of neural networks to learn low level interactions from properly developed descriptors. We hypothesize the use of spatially localized descriptors (i.e., the atomic environment vector[34] with modified angular symmetry function) within the cutoff to contribute greatly to this ability.

This contrasts with descriptor sets that represent the entire chemical environment at once, and thus interactions must be inferred through the entire set of non-local descriptors by the ML model.

Given the prospects of high-throughput experiments, robotic synthesis, and intelligent software, we are currently witnessing a transformation of science into a more data-driven automated discovery. The envisioned chemical AI imitates human decision making by transferring responsibility to an objective machine learning system. If successful overall, the approach will revolutionize the way computational methods are developed. As one possible building block to construct such AI, we introduced a fully automated workflow to select and calculate QM training data for accurate, transferable, and extensible ML potentials. These techniques can aid in the generation of universal potentials for a wide variety of current and future ML models.

## SUPPLEMENTARY INFORMATION

See supplementary material for complete technical details about training of ensemble of neural networks (S1.1) and sampling methods (S1.2). Tables S1-S10 list individual and complete COMP6 benchmarks. Table S11 lists details about ANI potential at various AL cycles. Tables S12-S17 list COMP6 benchmarks for ANI-1x within select energy ranges.

## ACKNOWLEDGEMENTS

J.S.S thanks the University of Florida for the graduate student fellowship and the Los Alamos National Laboratory Center for Non-linear Studies for resources and hospitality. This work was performed, in part, at the Center for Integrated Nanotechnologies, an Office of Science User Facility operated for the U.S. Department of Energy (DOE) Office of Science. Los Alamos National Laboratory, an affirmative action equal opportunity employer, is operated by Los Alamos National Security, LLC, for the National Nuclear Security Administration of the U.S. Department of Energy under contract DE-AC52-06NA25396. O.I. acknowledges support from DOD-ONR (N00014-16-1-2311) and Eshelman Institute for Innovation award. The authors acknowledge Extreme Science and Engineering Discovery Environment (XSEDE) award DMR110088, which is supported by National Science Foundation grant number ACI-1053575. This research in part was done using resources provided by the Open Science Grid[67,68] which is supported by the National Science Foundation award 1148698, and the U.S. Department of Energy's Office of Science. We gratefully acknowledge the support of the U.S. Department of Energy through the LANL/LDRD Program for this work. The authors thank Roman Zubatyuk and Kipton Barros for invaluable discussions on the topics presented in this work.

# Supplementary information for: "Less is more: sampling chemical space with active learning"

Justin S. Smith[1], Ben Nebgen[3], Nicholas Lubbers[3], Olexandr Isayev[2,*], Adrian E. Roitberg[1,*]

[1]Department of Chemistry, University of Florida, Gainesville, FL 32611, USA

[2]UNC Eshelman School of Pharmacy, University of North Carolina at Chapel Hill, Chapel Hill, NC 27599, USA

[3]Los Alamos National Laboratory, Los Alamos, NM 87545, USA

* Corresponding authors; email: OI (olexandr@olexandrisayev.com) and AER (roitberg@ufl.edu)

## S1 Methods

### S1.1 ANI Ensemble Preparation

Single network architectures vary by the size of the data set used to train the models during the active learning process. Table S11 describes all models presented in this work. Network sizes (depth and number of parameters) were determined through hyper parameter searches conducted at every configurational sampling step. Parameters for the atomic environment vector[1] (a numerical vector used to describe an atoms local chemical environment) used during the active learning process were constant and are provided with the released ANI-1x model. An initial learning rate of 0.001 is used. Early stopping is utilized in the training of each network, whereby if a model fails to improve its validation set predictions within 75 epochs then training is stopped. Learning rate annealing is utilized such that once a model stops early, training is restarted with a learning rate 0.5 times that of the previous learning rate. Termination of training is achieved when the learning rate is less than $1.0 \times 10^{-5}$. The adam[2] update method is used to update the weights during training.

Ensembles of ANI potentials are prepared using a 5-fold cross validation split of the data set and all previously mentioned hyper parameters. Training the ensemble to a 5-fold cross validation split ensures that the ensemble was trained to the entire data set for maximum performance. Rather than testing models on a 10% hold out from the training data set, we use benchmarks from the COMP6 benchmark suite to determine the fitness of an ensembles prediction. We do this because we are more interested in getting potentials which are not just accurate but also transferable and extensible. The COMP6 benchmarks provide more rigorous test case than could be achieved by testing on a 10% hold out of the training data set since molecules in the benchmarks are on average much larger than those included in the training set. This allows testing of extrapolation to larger structures, which is of great importance to any universal ML potential, rather than testing interpolation to already seen molecules. However, mean 10% hold out test set performance is supplied in Table S11 for the AL models published in this work.

### S1.2 Sampling methods

#### S1.2.1 Diverse Normal Mode Sampling (DNMS).

We modify the normal mode sampling (NMS) technique introduced by Smith el al.[1] to avoid data set clustering around equilibrium conformations. This technique identically follows NMS, in that a molecule is optimized at the desired QM level of theory, ωb97x with the 6-31g(d) basis set in this work, then frequency calculations are performed to obtain normal mode coordinates and their corresponding harmonic force constants. As with NMS, N random non-equilibrium conformations are generated by randomly perturbing the molecule along the normal mode coordinates. For diverse normal mode sampling (DNMS), the atomic environment vector[1] (referred to as AEV; a numerical vector used to describe the chemical environment of an atom in a molecule) for all C, N, and O atoms for each of the N conformations generated with NMS is stored. The squared Euclidean distance matrix between each of the N AEVs is computed. Finally, K diverse conformers is selected from the N original conformers using the max-min diversity selector algorithm implemented in the RDKit [http://www.rdkit.org/] cheminformatics software package.

For sampling, using the query by committee approach introduce in the main article we test all i conformers from the K selected diverse conformers. We generate QM energies and forces for all $\rho_i > \hat{\rho}$ (Section IIA of the main article) and add this new data to the training set in the next iteration of the active learning algorithm.

### S1.2.2 K Random Trajectory Sampling (KRTS).

Random trajectory sampling is carried out on a set of seed molecules from the conformational sampling data set. Given a molecular configuration, a random set of Boltzmann distributed velocities equal to 300K are generated. Molecular dynamics using the Langevin thermostat with 0.25fs time step at 300K is then initialized. The system is heated linearly over 4ps to 1000K. Every 5 steps of dynamics, $\rho_i$ (Section IIA of the main article) is computed. If $\rho_i > \hat{\rho}$ then dynamics is terminated. DFT reference data is then computed for the final conformation and added to the training set for the next iteration of the active learning cycle. If the trajectory reaches 4ps without encountering $\rho_i > \hat{\rho}$ then no new data is added to the training set. Many trajectories can be run for each of the seed molecules back to back to generate multiple new reference data.

### S1.2.3 MD Generated Dimer Sampling

Dimers are generated in an active learning scheme where a large box of hundreds of randomly selected small molecules (from the conformational sampling set) with random positions and orientation is generated. Molecular dynamics at 300K with periodic boundary conditions using the current version of the ANI active learned potential is ran on the box of molecules for X ps. After the molecular dynamics run, the box is decomposed into all dimers with intermolecular distances less than 5.0Å. Query by committee (Section S1.1) is then performed on all generated dimers, selecting any dimer with $\rho_i > \hat{\rho}$. DFT reference calculations are performed for all selected dimers to obtain refence training data. The new reference training data is added to the training data set for the next iteration of the active learning cycle. X was initially chosen to be small (10fs) but after time the algorithm stops generating new dimers, thus X is increased iteratively. As of the ANI-1x data set X is set to 5ps.

**Table S1.** *Complete COMP6 benchmark suite results for various ANI potentials.* Errors for conformer energy differences (ΔE), potential energies (E), and force components (F) for the active learned ANI potentials ANI-AL1 to ANI-1X compared with the original ANI-1 potential. These results are from the combination of all benchmarks within the COMP6 benchmark suite. µ and σ are the arithmetic mean and standard deviation, respectively. M and R are the MAE and RMSE, respectively. Units of energy are kcal $\times$ mol$^{-1}$ and units of force are kcal $\times$ mol$^{-1}$ $\times$ Å$^{-1}$.

| ANI Model | $\Delta E_M^\mu$ | $\Delta E_M^\sigma$ | $\Delta E_R^\mu$ | $\Delta E_R^\sigma$ | $E_M^\mu$ | $E_M^\sigma$ | $E_R^\mu$ | $E_R^\sigma$ | $F_M^\mu$ | $F_M^\sigma$ | $F_R^\mu$ | $F_R^\sigma$ |
|---|---|---|---|---|---|---|---|---|---|---|---|---|
| AL1 | 4.31 | 0.31 | 8.36 | 1.95 | 9.54 | 1.43 | 21.10 | 2.94 | 5.59 | 0.21 | 10.61 | 1.36 |
| AL2 | 3.56 | 0.47 | 6.42 | 1.88 | 10.21 | 1.46 | 24.80 | 6.52 | 5.19 | 0.57 | 9.72 | 2.03 |
| AL3 | 2.86 | 0.09 | 5.73 | 0.55 | 4.78 | 0.42 | 13.43 | 1.62 | 4.19 | 0.09 | 7.12 | 0.24 |
| AL4 | 2.72 | 0.09 | 5.32 | 0.30 | 5.04 | 0.50 | 14.40 | 1.79 | 4.01 | 0.10 | 6.89 | 0.41 |
| AL5 | 2.00 | 0.02 | 3.14 | 0.06 | 2.19 | 0.04 | 3.58 | 0.39 | 3.26 | 0.04 | 5.34 | 0.24 |
| AL6 | 1.85 | 0.04 | 2.95 | 0.15 | 1.93 | 0.08 | 3.37 | 0.78 | 3.09 | 0.09 | 5.29 | 0.58 |
| ANI-1 | 3.01 | 0.20 | 6.97 | 1.29 | 5.01 | 0.42 | 16.94 | 2.70 | 3.70 | 0.15 | 7.13 | 1.36 |

**Table S2.** *DrugBank COMP6 benchmark results for various ANI potentials.* Errors for conformer energy differences (ΔE), potential energies (E), and force components (F) for the active learned ANI potentials ANI-AL1 to ANI-1X compared with the original ANI-1 potential. µ and σ are the arithmetic mean and standard deviation, respectively. M and R are the MAE and RMSE, respectively. Units of energy are kcal $\times$ mol$^{-1}$ and units of force are kcal $\times$ mol$^{-1}$ $\times$ Å$^{-1}$.

| ANI Model | $\Delta E_M^\mu$ | $\Delta E_M^\sigma$ | $\Delta E_R^\mu$ | $\Delta E_R^\sigma$ | $E_M^\mu$ | $E_M^\sigma$ | $E_R^\mu$ | $E_R^\sigma$ | $F_M^\mu$ | $F_M^\sigma$ | $F_R^\mu$ | $F_R^\sigma$ |
|---|---|---|---|---|---|---|---|---|---|---|---|---|
| AL1 | 4.69 | 0.47 | 6.69 | 0.90 | 29.08 | 8.59 | 40.25 | 11.15 | 5.90 | 0.35 | 10.70 | 1.01 |
| AL2 | 4.94 | 1.63 | 7.16 | 2.92 | 30.35 | 8.71 | 42.80 | 14.31 | 5.97 | 1.47 | 11.17 | 4.16 |
| AL3 | 3.49 | 0.22 | 4.89 | 0.29 | 12.99 | 2.88 | 17.84 | 3.47 | 4.58 | 0.17 | 7.78 | 0.38 |
| AL4 | 3.51 | 0.30 | 4.89 | 0.53 | 14.78 | 2.23 | 19.33 | 2.96 | 4.49 | 0.34 | 7.84 | 1.20 |
| AL5 | 2.17 | 0.08 | 3.05 | 0.66 | 2.79 | 0.16 | 5.61 | 1.92 | 2.99 | 0.07 | 4.83 | 1.08 |
| AL6 | 2.09 | 0.14 | 3.18 | 0.83 | 2.65 | 0.28 | 6.01 | 3.01 | 2.86 | 0.16 | 5.35 | 1.82 |
| ANI-1 | 3.41 | 0.13 | 4.94 | 0.42 | 13.89 | 2.75 | 20.65 | 3.07 | 4.04 | 0.12 | 7.62 | 1.55 |

**Table S3.** *Tripeptide COMP6 benchmark.* Errors for conformer energy differences (ΔE), potential energies (E), and force components (F) for the active learned ANI potentials ANI-AL1 to ANI-1X compared with the original ANI-1 potential. µ and σ are the arithmetic mean and standard deviation, respectively. M and R are the MAE and RMSE, respectively. Units of energy are kcal $\times$ mol$^{-1}$ and units of force are kcal $\times$ mol$^{-1}$ $\times$ Å$^{-1}$.

| ANI Model | $\Delta E_M^\mu$ | $\Delta E_M^\sigma$ | $\Delta E_R^\mu$ | $\Delta E_R^\sigma$ | $E_M^\mu$ | $E_M^\sigma$ | $E_R^\mu$ | $E_R^\sigma$ | $F_M^\mu$ | $F_M^\sigma$ | $F_R^\mu$ | $F_R^\sigma$ |
|---|---|---|---|---|---|---|---|---|---|---|---|---|
| AL1 | 4.75 | 0.36 | 13.79 | 2.25 | 15.25 | 3.29 | 21.08 | 3.94 | 4.79 | 0.21 | 11.45 | 0.86 |
| AL2 | 4.08 | 0.76 | 10.02 | 1.50 | 20.92 | 4.70 | 27.82 | 7.78 | 4.90 | 0.89 | 11.04 | 1.87 |
| AL3 | 3.54 | 0.10 | 8.75 | 0.96 | 11.56 | 2.27 | 15.72 | 2.38 | 4.26 | 0.14 | 9.18 | 0.25 |
| AL4 | 2.92 | 0.12 | 7.22 | 1.07 | 13.17 | 3.33 | 15.35 | 3.00 | 3.95 | 0.14 | 7.77 | 0.73 |
| AL5 | 1.78 | 0.05 | 2.90 | 0.30 | 3.50 | 0.10 | 4.51 | 0.16 | 2.67 | 0.05 | 4.43 | 1.80 |
| AL6 | 1.65 | 0.07 | 2.58 | 0.48 | 2.92 | 0.22 | 3.77 | 0.47 | 2.49 | 0.04 | 4.79 | 0.70 |
| ANI-1 | 2.86 | 0.31 | 4.99 | 0.69 | 13.10 | 1.15 | 16.47 | 1.70 | 3.46 | 0.34 | 6.67 | 1.58 |

**Table S4.** *GDB07to09 COMP6 benchmark.* Errors for conformer energy differences (ΔE), potential energies (E), and force components (F) for the active learned ANI potentials ANI-AL1 to ANI-1X compared with the original ANI-1 potential. μ and σ are the arithmetic mean and standard deviation, respectively. M and R are the MAE and RMSE, respectively. Units of energy are kcal $\times$ mol$^{-1}$ and units of force are kcal $\times$ mol$^{-1}$ $\times$ Å$^{-1}$.

| ANI Model | $\Delta E_M^\mu$ | $\Delta E_M^\sigma$ | $\Delta E_R^\mu$ | $\Delta E_R^\sigma$ | $E_M^\mu$ | $E_M^\sigma$ | $E_R^\mu$ | $E_R^\sigma$ | $F_M^\mu$ | $F_M^\sigma$ | $F_R^\mu$ | $F_R^\sigma$ |
|---|---|---|---|---|---|---|---|---|---|---|---|---|
| **AL1** | 2.14 | 0.03 | 4.25 | 0.11 | 2.99 | 0.12 | 5.72 | 0.35 | 3.92 | 0.03 | 7.76 | 0.57 |
| **AL2** | 1.66 | 0.03 | 2.83 | 0.13 | 2.82 | 0.09 | 5.53 | 0.34 | 3.39 | 0.08 | 6.35 | 0.42 |
| **AL3** | 1.33 | 0.02 | 2.09 | 0.14 | 1.43 | 0.02 | 2.18 | 0.29 | 2.95 | 0.03 | 4.94 | 0.25 |
| **AL4** | 1.26 | 0.01 | 1.92 | 0.05 | 1.34 | 0.04 | 1.89 | 0.06 | 2.81 | 0.02 | 4.59 | 0.06 |
| **AL5** | 1.18 | 0.02 | 1.79 | 0.02 | 1.20 | 0.02 | 1.69 | 0.02 | 2.64 | 0.04 | 4.24 | 0.04 |
| **AL6** | 1.07 | 0.03 | 1.66 | 0.05 | 1.04 | 0.04 | 1.50 | 0.05 | 2.43 | 0.07 | 3.93 | 0.08 |
| **ANI-1** | 1.28 | 0.01 | 2.19 | 0.05 | 1.30 | 0.01 | 2.31 | 0.04 | 2.50 | 0.03 | 4.57 | 0.16 |

**Table S5.** *GDB10to13 COMP6 benchmark.* Errors for conformer energy differences (ΔE), potential energies (E), and force components (F) for the active learned ANI potentials ANI-AL1 to ANI-1X compared with the original ANI-1 potential. μ and σ are the arithmetic mean and standard deviation, respectively. M and R are the MAE and RMSE, respectively. Units of energy are kcal $\times$ mol$^{-1}$ and units of force are kcal $\times$ mol$^{-1}$ $\times$ Å$^{-1}$.

| ANI Model | $\Delta E_M^\mu$ | $\Delta E_M^\sigma$ | $\Delta E_R^\mu$ | $\Delta E_R^\sigma$ | $E_M^\mu$ | $E_M^\sigma$ | $E_R^\mu$ | $E_R^\sigma$ | $F_M^\mu$ | $F_M^\sigma$ | $F_R^\mu$ | $F_R^\sigma$ |
|---|---|---|---|---|---|---|---|---|---|---|---|---|
| **AL1** | 5.25 | 0.14 | 8.01 | 0.30 | 7.18 | 0.43 | 11.00 | 0.60 | 6.38 | 0.19 | 11.72 | 0.69 |
| **AL2** | 4.12 | 0.24 | 6.25 | 0.52 | 7.70 | 0.98 | 13.43 | 2.38 | 5.65 | 0.39 | 10.03 | 1.10 |
| **AL3** | 3.13 | 0.05 | 4.60 | 0.14 | 3.29 | 0.10 | 4.69 | 0.25 | 4.56 | 0.09 | 7.42 | 0.26 |
| **AL4** | 2.89 | 0.04 | 4.19 | 0.08 | 3.32 | 0.21 | 4.48 | 0.27 | 4.31 | 0.06 | 7.03 | 0.16 |
| **AL5** | 2.57 | 0.04 | 3.70 | 0.06 | 2.62 | 0.05 | 3.58 | 0.05 | 3.85 | 0.05 | 6.23 | 0.08 |
| **AL6** | 2.38 | 0.05 | 3.47 | 0.11 | 2.30 | 0.06 | 3.21 | 0.14 | 3.67 | 0.09 | 6.01 | 0.17 |
| **ANI-1** | 2.98 | 0.05 | 4.54 | 0.16 | 3.12 | 0.11 | 4.74 | 0.24 | 3.96 | 0.06 | 7.09 | 0.24 |

**Table S6.** *S66x8 COMP6 benchmark.* Errors for conformer energy differences (ΔE), potential energies (E), and force components (F) for the active learned ANI potentials ANI-AL1 to ANI-1X compared with the original ANI-1 potential. μ and σ are the arithmetic mean and standard deviation, respectively. M and R are the MAE and RMSE, respectively. Units of energy are kcal $\times$ mol$^{-1}$ and units of force are kcal $\times$ mol$^{-1}$ $\times$ Å$^{-1}$.

| ANI Model | $\Delta E_M^\mu$ | $\Delta E_M^\sigma$ | $\Delta E_R^\mu$ | $\Delta E_R^\sigma$ | $E_M^\mu$ | $E_M^\sigma$ | $E_R^\mu$ | $E_R^\sigma$ | $F_M^\mu$ | $F_M^\sigma$ | $F_R^\mu$ | $F_R^\sigma$ |
|---|---|---|---|---|---|---|---|---|---|---|---|---|
| **AL1** | 8.05 | 0.73 | 12.28 | 1.25 | 11.63 | 0.34 | 16.47 | 0.44 | 2.67 | 0.12 | 5.77 | 0.21 |
| **AL2** | 8.98 | 1.40 | 13.77 | 2.37 | 11.95 | 1.60 | 17.09 | 2.47 | 2.63 | 0.31 | 5.87 | 2.58 |
| **AL3** | 8.18 | 0.84 | 12.95 | 1.49 | 9.99 | 1.01 | 14.99 | 1.57 | 2.37 | 0.14 | 4.78 | 0.82 |
| **AL4** | 6.55 | 0.29 | 10.85 | 0.71 | 8.54 | 0.52 | 12.71 | 0.74 | 2.41 | 0.23 | 4.54 | 0.39 |
| **AL5** | 1.57 | 0.08 | 2.26 | 0.11 | 2.51 | 0.17 | 3.45 | 0.21 | 1.72 | 0.06 | 3.07 | 0.17 |
| **AL6** | 1.42 | 0.09 | 2.10 | 0.26 | 2.06 | 0.13 | 3.01 | 0.18 | 1.60 | 0.15 | 2.76 | 0.30 |
| **ANI-1** | 10.32 | 0.76 | 16.76 | 1.07 | 13.25 | 0.96 | 20.12 | 1.38 | 3.17 | 0.36 | 9.08 | 4.07 |

**Table S7. *ANI-MD COMP6 benchmark.*** Errors for conformer energy differences (ΔE), potential energies (E), and force components (F) for the active learned ANI potentials AL1 to AL6 compared with the original ANI-1 potential. μ and σ are the arithmetic mean and standard deviation, respectively. M and R are the MAE and RMSE, respectively. Units of energy are kcal $\times$ mol$^{-1}$ and units of force are kcal $\times$ mol$^{-1} \times$ Å$^{-1}$.

| ANI Model | $\Delta E_M^\mu$ | $\Delta E_M^\sigma$ | $\Delta E_R^\mu$ | $\Delta E_R^\sigma$ | $E_M^\mu$ | $E_M^\sigma$ | $E_R^\mu$ | $E_R^\sigma$ | $F_M^\mu$ | $F_M^\sigma$ | $F_R^\mu$ | $F_R^\sigma$ |
|---|---|---|---|---|---|---|---|---|---|---|---|---|
| **AL1** | 8.47 | 2.34 | 17.11 | 7.89 | 51.55 | 6.19 | 92.95 | 12.64 | 5.95 | 1.47 | 11.11 | 10.58 |
| **AL2** | 7.15 | 2.76 | 12.46 | 6.76 | 62.92 | 20.70 | 121.5 | 55.94 | 6.37 | 3.02 | 11.70 | 10.29 |
| **AL3** | 6.80 | 0.76 | 13.43 | 2.33 | 41.41 | 5.10 | 82.56 | 13.44 | 4.99 | 0.37 | 8.11 | 1.17 |
| **AL4** | 6.61 | 0.53 | 12.52 | 1.01 | 42.41 | 4.82 | 89.62 | 12.82 | 5.01 | 0.28 | 8.99 | 1.54 |
| **AL5** | 2.82 | 0.09 | 4.61 | 0.28 | 4.39 | 0.73 | 8.10 | 1.88 | 2.89 | 0.06 | 4.47 | 0.10 |
| **AL6** | 2.59 | 0.10 | 4.17 | 0.26 | 3.40 | 0.65 | 5.94 | 1.48 | 2.68 | 0.16 | 4.24 | 0.63 |
| **ANI-1** | 8.93 | 1.58 | 18.08 | 4.39 | 52.30 | 8.84 | 109.2 | 25.32 | 5.80 | 1.61 | 12.70 | 7.80 |

**Table S8. *Individual ANI-MD COMP6 benchmark trajectories.*** Errors for conformer energy differences (ΔE), potential energies (E), and force components (F) for the ANI-1x potential vs DFT reference calculations on the 128 conformations per molecule in the ANI-MD COMP6 benchmark. Units of energy are kcal $\times$ mol$^{-1}$ and units of force are kcal $\times$ mol$^{-1} \times$ Å$^{-1}$. Per conformation (conf.) energy and force prediction timings are also included for the ANI-1x potential.

| System | E MAE | E RMSE | E range | F MAE | F RMSE | F range | ΔE MAE | ΔE RMSE | ΔE range | Time(ms) per conf. |
|---|---|---|---|---|---|---|---|---|---|---|
| **Acetaminophen** | 0.56 | 0.70 | 15.7 | 2.06 | 2.96 | 196.6 | 0.80 | 1.00 | 30.2 | 2.7 |
| **Caffeine** | 0.98 | 1.24 | 17.1 | 3.56 | 5.46 | 260.1 | 1.31 | 1.66 | 34.1 | 3.0 |
| **Salbutamol** | 1.81 | 2.10 | 21.2 | 2.16 | 3.05 | 226.8 | 1.23 | 1.55 | 42.1 | 4.0 |
| **Atomoxetine** | 1.19 | 1.50 | 19.7 | 1.84 | 2.62 | 203.0 | 1.51 | 1.90 | 36.5 | 4.3 |
| **Lisdexamfetamine** | 0.94 | 1.18 | 27.9 | 1.63 | 2.27 | 229.5 | 1.29 | 1.66 | 54.4 | 3.8 |
| **Capsaicin** | 1.53 | 1.98 | 25.8 | 2.01 | 2.91 | 231.1 | 2.07 | 2.61 | 50.3 | 3.6 |
| **Oseltamivir** | 2.96 | 3.45 | 40.7 | 2.41 | 3.68 | 233.5 | 1.95 | 2.57 | 79.7 | 4.5 |
| **Retinol** | 3.68 | 4.72 | 41.0 | 3.88 | 7.70 | 285.6 | 3.99 | 5.08 | 75.5 | 4.0 |
| **Fentanyl** | 1.26 | 1.58 | 29.8 | 2.03 | 2.92 | 208.1 | 1.69 | 2.13 | 55.3 | 4.3 |
| **Tolterodine** | 1.91 | 2.28 | 28.3 | 2.10 | 3.04 | 205.1 | 1.66 | 2.07 | 54.4 | 5.3 |
| **Ranolazine** | 2.22 | 2.62 | 29.1 | 2.20 | 3.06 | 237.1 | 1.95 | 2.45 | 57.2 | 4.4 |
| **Atazanavir** | 2.73 | 3.52 | 52.0 | 2.53 | 3.80 | 242.9 | 3.92 | 4.94 | 101.1 | 5.0 |
| **Chignolin (1UAO)** | 18.07 | 18.47 | 45.5 | 3.25 | 4.68 | 325.3 | 4.36 | 5.42 | 86.2 | 5.5 |
| **TrpCage (1L2Y)** | 16.34 | 18.37 | 102.5 | 3.13 | 4.59 | 249.7 | 9.80 | 12.10 | 191.0 | 9.3 |

**Table S9.** *Individual ANI-MD COMP6 benchmark trajectories per atom.* Per atom errors for the conformer energy differences ($\Delta E$) and potential energies ($E$) for the ANI-1x potential vs DFT reference calculations on the 128 conformations per molecule in the ANI-MD COMP6 benchmark. Units of energy are kcal $\times$ mol$^{-1}$.

| System | # of atoms | $\frac{E_{RMS}}{\sqrt{N}}$ | $\frac{E_{RMS}}{N}$ | $\frac{\Delta E_{RMS}}{\sqrt{N}}$ | $\frac{\Delta E_{RMS}}{N}$ |
|---|---|---|---|---|---|
| **Acetaminophen** | 20 | 0.16 | 0.04 | 0.22 | 0.05 |
| **Caffeine** | 24 | 0.25 | 0.05 | 0.34 | 0.07 |
| **Salbutamol** | 38 | 0.34 | 0.06 | 0.25 | 0.04 |
| **Atomoxetine** | 40 | 0.24 | 0.04 | 0.30 | 0.05 |
| **Lisdexamfetamine** | 44 | 0.18 | 0.03 | 0.25 | 0.04 |
| **Capsaicin** | 49 | 0.28 | 0.04 | 0.37 | 0.05 |
| **Oseltamivir** | 50 | 0.49 | 0.07 | 0.36 | 0.05 |
| **Retinol** | 51 | 0.66 | 0.09 | 0.71 | 0.10 |
| **Fentanyl** | 53 | 0.22 | 0.03 | 0.29 | 0.04 |
| **Tolterodine** | 55 | 0.31 | 0.04 | 0.28 | 0.04 |
| **Ranolazine** | 64 | 0.33 | 0.04 | 0.31 | 0.04 |
| **Atazanavir** | 103 | 0.35 | 0.03 | 0.49 | 0.05 |
| **Chignolin (1UAO)** | 149 | 1.51 | 0.12 | 0.44 | 0.04 |
| **TrpCage (1L2Y)** | 312 | 1.04 | 0.06 | 0.69 | 0.04 |
| **Mean** | **75** | **0.45** | **0.05** | **0.38** | **0.05** |

**Table S10.** *Individual ANI-MD trajectories COMP6 benchmark for ANI, DFTB, and PM6.* ANI-1x, DFTB and PM6 vs DFT reference calculation errors for the conformer energy differences ($\Delta E$) and force components (F) on the 128 conformations per molecule in the ANI-MD COMP6 benchmark. Units of energy are kcal $\times$ mol$^{-1}$ and units of force are kcal $\times$ mol$^{-1}$ $\times$ Å$^{-1}$.

| System | $F_{MAE}^{ANI}$ | $F_{MAE}^{DFTB}$ | $F_{MAE}^{PM6}$ | $F_{RMS}^{ANI}$ | $F_{RMS}^{DFTB}$ | $F_{RMS}^{PM6}$ | $\Delta E_{MAE}^{ANI}$ | $\Delta E_{MAE}^{DFTB}$ | $\Delta E_{MAE}^{PM6}$ | $\Delta E_{RMS}^{ANI}$ | $\Delta E_{RMS}^{DFTB}$ | $\Delta E_{RMS}^{PM6}$ |
|---|---|---|---|---|---|---|---|---|---|---|---|---|
| **Acetaminophen** | 2.06 | 4.87 | 7.23 | 2.96 | 7.16 | 10.8 | 0.80 | 1.59 | 2.23 | 1.00 | 2.00 | 2.80 |
| **Atazanavir** | 2.53 | 5.11 | 8.14 | 3.80 | 7.65 | 12.1 | 3.92 | 4.63 | 7.43 | 4.94 | 5.71 | 9.27 |
| **Atomoxetine** | 1.84 | 3.23 | 6.38 | 2.62 | 4.57 | 9.25 | 1.51 | 2.41 | 3.12 | 1.90 | 3.02 | 3.85 |
| **Caffeine** | 3.56 | 7.30 | 12.0 | 5.46 | 12.1 | 19.8 | 1.31 | 2.33 | 4.92 | 1.66 | 2.93 | 6.08 |
| **Capsaicin** | 2.01 | 3.58 | 6.72 | 2.91 | 5.33 | 9.86 | 2.07 | 2.43 | 3.76 | 2.61 | 3.02 | 4.67 |
| **Chignolin (1UAO)** | 3.25 | 5.87 | 9.76 | 4.68 | 8.44 | 13.7 | 4.36 | 5.50 | 9.62 | 5.42 | 7.05 | 12.1 |
| **Fentanyl** | 2.03 | 3.46 | 6.71 | 2.92 | 5.12 | 9.60 | 1.69 | 1.98 | 3.81 | 2.13 | 2.47 | 4.76 |
| **Lisdexamfetamine** | 1.63 | 3.53 | 6.84 | 2.27 | 5.26 | 9.89 | 1.29 | 2.27 | 4.24 | 1.66 | 2.84 | 5.29 |
| **Oseltamivir** | 2.41 | 4.56 | 7.31 | 3.68 | 6.79 | 10.4 | 1.95 | 2.76 | 4.00 | 2.57 | 3.47 | 4.94 |
| **Ranolazine** | 2.20 | 4.20 | 8.26 | 3.06 | 5.92 | 11.7 | 1.95 | 4.01 | 6.03 | 2.45 | 4.97 | 7.49 |
| **Retinol** | 3.88 | 4.31 | 5.21 | 7.70 | 6.99 | 7.16 | 3.99 | 2.51 | 3.05 | 5.08 | 3.16 | 3.85 |
| **Salbutamol** | 2.16 | 4.20 | 7.28 | 3.05 | 5.46 | 10.2 | 1.23 | 2.46 | 3.56 | 1.55 | 3.05 | 4.46 |
| **Tolterodine** | 2.10 | 3.54 | 5.92 | 3.04 | 4.89 | 8.29 | 1.66 | 2.55 | 3.44 | 2.07 | 3.16 | 4.36 |
| **TrpCage (1L2Y)** | 3.13 | 5.38 | 9.07 | 4.59 | 7.87 | 12.7 | 9.80 | 9.84 | 13.1 | 12.1 | 12.3 | 16.3 |
| **Mean** | **2.49** | **4.51** | **7.63** | **3.77** | **6.68** | **11.1** | **2.68** | **3.38** | **5.16** | **3.37** | **4.23** | **6.45** |

**Table S11.** *ANI model details.* Model details on the individual active learned ANI models (ANI-AL) and a network ensemble trained to the original ANI-1 data set. AL version is an internal versioning scheme which allows tracking of the data included at any given model version. AL cycles are the number of active learning conformational search cycles completed to produce the given ANI-AL model. Parameters is the total number of parameters in the models, all of which consisted of an input size of 386 and 3 total hidden layers with varying numbers of nodes per layer. Test set potential energy (E) RMSE are provided in $\text{kcal} \times \text{mol}^{-1}$.

| Model | AL Cycles | Parameters | Test set RMSE (E) | Configurational Sampling |
|---|---|---|---|---|
| AL1 | 0 | 57472 | 0.8 | GDB-1 to 6 |
| AL2 | 9 | 84096 | 1.1 | GDB-1 to 6 |
| AL3 | 13 | 84096 | 1.6 | GDB-1 to 7 |
| AL4 | 18 | 120000 | 2.2 | GDB-1 to 8 |
| AL5 | 32 | 184512 | 2.6 | GDB-1 to 8; GDB-1 to 6 dimers; amino acids; dipeptides; ChEMBL |
| ANI-1X (AL6) | 37 | 389376 | 2.7 | GDB-1 to 9; GDB-1 to 6 dimers; amino acids; dipeptides; ChMBL |
| ANI-1 | - | 270592 | 1.2 | GDB-1 to 8 |

**Table S12.** *Complete COMP6 benchmark for ANI-1x within select energy ranges.* Errors for conformer energy differences (ΔE), potential energies (E), and force components (F) for the active learned ANI potential ANI-1x over select energy ranges of the test set. The test set for a given energy range is built by only considering conformations of a given molecule within the energy range (shown in column 2) from the minimum energy conformer in the set of conformations. These results are from the combination of all benchmarks within the COMP6 benchmark suite. μ and σ are the arithmetic mean and standard deviation, respectively. M and R are the MAE and RMSE, respectively. Units of energy are $\text{kcal} \times \text{mol}^{-1}$ and units of force are $\text{kcal} \times \text{mol}^{-1} \times \text{Å}^{-1}$.

| ANI Model | Energy Range | $\Delta E_M^\mu$ | $\Delta E_M^\sigma$ | $\Delta E_R^\mu$ | $\Delta E_R^\sigma$ | $E_M^\mu$ | $E_M^\sigma$ | $E_R^\mu$ | $E_R^\sigma$ | $F_M^\mu$ | $F_M^\sigma$ | $F_R^\mu$ | $F_R^\sigma$ |
|---|---|---|---|---|---|---|---|---|---|---|---|---|---|
| ANI-1x | 10 | 0.65 | 0.01 | 1.00 | 0.05 | 1.29 | 0.06 | 2.60 | 0.93 | 2.21 | 0.08 | 3.68 | 0.49 |
|  | 30 | 1.05 | 0.02 | 1.65 | 0.06 | 1.35 | 0.06 | 2.53 | 0.70 | 2.35 | 0.08 | 3.84 | 0.41 |
|  | 50 | 1.19 | 0.03 | 1.97 | 0.08 | 1.44 | 0.07 | 2.71 | 0.81 | 2.48 | 0.08 | 4.11 | 0.56 |
|  | 100 | 1.39 | 0.03 | 2.28 | 0.11 | 1.61 | 0.07 | 3.01 | 0.91 | 2.70 | 0.09 | 4.53 | 0.65 |
|  | 150 | 1.58 | 0.04 | 2.54 | 0.13 | 1.75 | 0.08 | 3.19 | 0.88 | 2.87 | 0.09 | 4.85 | 0.64 |
|  | 200 | 1.71 | 0.04 | 2.73 | 0.13 | 1.84 | 0.08 | 3.28 | 0.84 | 2.98 | 0.09 | 5.06 | 0.60 |
|  | 250 | 1.79 | 0.04 | 2.85 | 0.15 | 1.89 | 0.08 | 3.33 | 0.81 | 3.04 | 0.10 | 5.18 | 0.58 |
|  | 300 | 1.82 | 0.04 | 2.91 | 0.15 | 1.91 | 0.08 | 3.35 | 0.79 | 3.07 | 0.10 | 5.23 | 0.57 |

**Table S13.** *DrugBank COMP6 benchmark for ANI-1x within select energy ranges.* Errors for conformer energy differences (ΔE), potential energies (E), and force components (F) for the active learned ANI potential ANI-1x over select energy ranges of the test set. The test set for a given energy range is built by only considering conformations of a given molecule within the energy range (shown in column 2) from the minimum energy conformer in the set of conformations. μ and σ are the arithmetic mean and standard deviation, respectively. M and R are the MAE and RMSE, respectively. Units of energy are kcal $\times$ mol$^{-1}$ and units of force are kcal $\times$ mol$^{-1}$ $\times$ Å$^{-1}$.

| ANI Model | Energy Range | $\Delta E_M^\mu$ | $\Delta E_M^\sigma$ | $\Delta E_R^\mu$ | $\Delta E_R^\sigma$ | $E_M^\mu$ | $E_M^\sigma$ | $E_R^\mu$ | $E_R^\sigma$ | $F_M^\mu$ | $F_M^\sigma$ | $F_R^\mu$ | $F_R^\sigma$ |
|---|---|---|---|---|---|---|---|---|---|---|---|---|---|
| ANI-1x | 10 | 0.70 | 0.10 | 1.33 | 0.67 | 2.22 | 0.36 | 6.83 | 4.25 | 2.30 | 0.19 | 4.75 | 1.98 |
| | 30 | 0.92 | 0.08 | 1.58 | 0.66 | 2.11 | 0.27 | 5.82 | 3.37 | 2.36 | 0.16 | 4.73 | 1.74 |
| | 50 | 1.16 | 0.13 | 2.08 | 1.02 | 2.22 | 0.31 | 6.06 | 3.57 | 2.48 | 0.18 | 5.20 | 2.23 |
| | 100 | 1.58 | 0.13 | 2.61 | 1.05 | 2.38 | 0.31 | 6.15 | 3.45 | 2.66 | 0.18 | 5.37 | 2.24 |
| | 150 | 1.83 | 0.14 | 2.89 | 1.00 | 2.51 | 0.30 | 6.12 | 3.28 | 2.77 | 0.17 | 5.35 | 2.09 |
| | 200 | 1.97 | 0.14 | 3.04 | 0.92 | 2.59 | 0.29 | 6.05 | 3.13 | 2.82 | 0.17 | 5.37 | 1.95 |
| | 250 | 2.04 | 0.14 | 3.11 | 0.87 | 2.62 | 0.28 | 6.02 | 3.05 | 2.85 | 0.16 | 5.36 | 1.87 |
| | 300 | 2.07 | 0.14 | 3.15 | 0.85 | 2.64 | 0.28 | 6.01 | 3.02 | 2.86 | 0.16 | 5.35 | 1.84 |

**Table S14.** *Tripeptide COMP6 benchmark for ANI-1x within select energy ranges.* Errors for conformer energy differences (ΔE), potential energies (E), and force components (F) for the active learned ANI potential ANI-1x over select energy ranges of the test set. The test set for a given energy range is built by only considering conformations of a given molecule within the energy range (shown in column 2) from the minimum energy conformer in the set of conformations. μ and σ are the arithmetic mean and standard deviation, respectively. M and R are the MAE and RMSE, respectively. Units of energy are kcal $\times$ mol$^{-1}$ and units of force are kcal $\times$ mol$^{-1}$ $\times$ Å$^{-1}$.

| ANI Model | Energy Range | $\Delta E_M^\mu$ | $\Delta E_M^\sigma$ | $\Delta E_R^\mu$ | $\Delta E_R^\sigma$ | $E_M^\mu$ | $E_M^\sigma$ | $E_R^\mu$ | $E_R^\sigma$ | $F_M^\mu$ | $F_M^\sigma$ | $F_R^\mu$ | $F_R^\sigma$ |
|---|---|---|---|---|---|---|---|---|---|---|---|---|---|
| ANI-1x | 10 | 0.64 | 0.03 | 0.91 | 0.05 | 2.48 | 0.24 | 3.05 | 0.25 | 2.12 | 0.04 | 3.16 | 0.06 |
| | 30 | 0.90 | 0.02 | 1.20 | 0.03 | 2.59 | 0.25 | 3.20 | 0.26 | 2.18 | 0.04 | 3.24 | 0.06 |
| | 50 | 1.07 | 0.02 | 1.43 | 0.04 | 2.67 | 0.23 | 3.28 | 0.24 | 2.25 | 0.04 | 3.33 | 0.06 |
| | 100 | 1.35 | 0.02 | 1.80 | 0.04 | 2.75 | 0.20 | 3.38 | 0.21 | 2.36 | 0.03 | 3.53 | 0.06 |
| | 150 | 1.48 | 0.04 | 2.00 | 0.07 | 2.81 | 0.19 | 3.45 | 0.18 | 2.41 | 0.04 | 3.72 | 0.09 |
| | 200 | 1.53 | 0.05 | 2.09 | 0.10 | 2.84 | 0.19 | 3.51 | 0.18 | 2.44 | 0.04 | 3.87 | 0.17 |
| | 250 | 1.56 | 0.05 | 2.19 | 0.11 | 2.86 | 0.19 | 3.58 | 0.19 | 2.45 | 0.04 | 3.98 | 0.18 |
| | 300 | 1.59 | 0.05 | 2.34 | 0.11 | 2.88 | 0.19 | 3.64 | 0.18 | 2.46 | 0.04 | 4.21 | 0.28 |

**Table S15.** *GDB07to09 COMP6 benchmark for ANI-1x within select energy ranges.* Errors for conformer energy differences (ΔE), potential energies (E), and force components (F) for the active learned ANI potential ANI-1x over select energy ranges of the test set. The test set for a given energy range is built by only considering conformations of a given molecule within the energy range (shown in column 2) from the minimum energy conformer in the set of conformations. μ and σ are the arithmetic mean and standard deviation, respectively. M and R are the MAE and RMSE, respectively. Units of energy are kcal $\times$ mol$^{-1}$ and units of force are kcal $\times$ mol$^{-1}$ $\times$ Å$^{-1}$.

| ANI Model | Energy Range | $\Delta E_M^\mu$ | $\Delta E_M^\sigma$ | $\Delta E_R^\mu$ | $\Delta E_R^\sigma$ | $E_M^\mu$ | $E_M^\sigma$ | $E_R^\mu$ | $E_R^\sigma$ | $F_M^\mu$ | $F_M^\sigma$ | $F_R^\mu$ | $F_R^\sigma$ |
|---|---|---|---|---|---|---|---|---|---|---|---|---|---|
| ANI-1x | 10 | 0.46 | 0.01 | 0.62 | 0.02 | 0.80 | 0.04 | 1.09 | 0.05 | 1.93 | 0.06 | 2.96 | 0.08 |
| | 30 | 0.71 | 0.02 | 0.95 | 0.02 | 0.88 | 0.04 | 1.18 | 0.05 | 2.12 | 0.06 | 3.22 | 0.09 |
| | 50 | 0.84 | 0.02 | 1.15 | 0.03 | 0.93 | 0.04 | 1.25 | 0.05 | 2.24 | 0.06 | 3.42 | 0.09 |
| | 100 | 0.99 | 0.03 | 1.41 | 0.04 | 1.00 | 0.04 | 1.37 | 0.05 | 2.37 | 0.07 | 3.71 | 0.10 |
| | 150 | 1.05 | 0.03 | 1.58 | 0.04 | 1.03 | 0.04 | 1.46 | 0.05 | 2.42 | 0.07 | 3.87 | 0.10 |
| | 200 | 1.07 | 0.03 | 1.65 | 0.05 | 1.04 | 0.04 | 1.49 | 0.05 | 2.43 | 0.07 | 3.92 | 0.08 |
| | 250 | 1.07 | 0.03 | 1.66 | 0.05 | 1.04 | 0.04 | 1.50 | 0.05 | 2.43 | 0.07 | 3.93 | 0.08 |
| | 300 | 1.07 | 0.03 | 1.66 | 0.05 | 1.04 | 0.04 | 1.50 | 0.05 | 2.43 | 0.07 | 3.93 | 0.08 |

**Table S16.** *GDB10to13 COMP6 benchmark for ANI-1x within select energy ranges.* Errors for conformer energy differences (ΔE), potential energies (E), and force components (F) for the active learned ANI potential ANI-1x over select energy ranges of the test set. The test set for a given energy range is built by only considering conformations of a given molecule within the energy range (shown in column 2) from the minimum energy conformer in the set of conformations. μ and σ are the arithmetic mean and standard deviation, respectively. M and R are the MAE and RMSE, respectively. Units of energy are kcal $\times$ mol$^{-1}$ and units of force are kcal $\times$ mol$^{-1}$ $\times$ Å$^{-1}$.

| ANI Model | Energy Range | $\Delta E_M^\mu$ | $\Delta E_M^\sigma$ | $\Delta E_R^\mu$ | $\Delta E_R^\sigma$ | $E_M^\mu$ | $E_M^\sigma$ | $E_R^\mu$ | $E_R^\sigma$ | $F_M^\mu$ | $F_M^\sigma$ | $F_R^\mu$ | $F_R^\sigma$ |
|---|---|---|---|---|---|---|---|---|---|---|---|---|---|
| ANI-1x | 10 | 0.61 | 0.01 | 0.86 | 0.02 | 1.63 | 0.03 | 2.18 | 0.04 | 2.48 | 0.07 | 3.94 | 0.12 |
| | 30 | 0.90 | 0.02 | 1.24 | 0.03 | 1.66 | 0.03 | 2.26 | 0.09 | 2.61 | 0.07 | 4.13 | 0.12 |
| | 50 | 1.09 | 0.02 | 1.51 | 0.03 | 1.72 | 0.03 | 2.34 | 0.08 | 2.75 | 0.07 | 4.34 | 0.12 |
| | 100 | 1.47 | 0.03 | 2.12 | 0.06 | 1.88 | 0.04 | 2.60 | 0.11 | 3.04 | 0.08 | 4.85 | 0.13 |
| | 150 | 1.86 | 0.04 | 2.70 | 0.06 | 2.05 | 0.05 | 2.85 | 0.10 | 3.32 | 0.08 | 5.35 | 0.14 |
| | 200 | 2.13 | 0.04 | 3.09 | 0.08 | 2.18 | 0.05 | 3.03 | 0.11 | 3.50 | 0.09 | 5.67 | 0.16 |
| | 250 | 2.27 | 0.05 | 3.30 | 0.11 | 2.24 | 0.05 | 3.13 | 0.14 | 3.60 | 0.09 | 5.87 | 0.17 |
| | 300 | 2.34 | 0.05 | 3.40 | 0.11 | 2.27 | 0.05 | 3.18 | 0.14 | 3.64 | 0.09 | 5.95 | 0.17 |

**Table S17.** *Complete COMP6 benchmark per atom errors for ANI-1x within select energy ranges.* Per atom errors for conformer energy differences ($\Delta E$) and potential energies ($E$) achieved by the active learned ANI potential ANI-1x over select energy ranges of the entire COMP6 benchmark. The test set for a given energy range is built by only considering conformations of a given molecule within the energy range (shown in column 2) from the minimum energy conformer in the set of conformations. μ is the arithmetic mean. M and R are the MAE and RMSE, respectively. Units of energy are kcal × mol$^{-1}$ per atom.

| ANI Model | Energy Range | $\Delta E_M^\mu$ | $\Delta E_R^\mu$ | $E_M^\mu$ | $E_R^\mu$ |
|---|---|---|---|---|---|
| ANI-1x | 10 | 0.031 | 0.044 | 0.059 | 0.101 |
| | 30 | 0.042 | 0.058 | 0.061 | 0.099 |
| | 50 | 0.049 | 0.070 | 0.064 | 0.105 |
| | 100 | 0.059 | 0.090 | 0.070 | 0.117 |
| | 150 | 0.067 | 0.107 | 0.075 | 0.125 |
| | 200 | 0.072 | 0.116 | 0.077 | 0.129 |
| | 250 | 0.074 | 0.120 | 0.078 | 0.130 |
| | 300 | 0.075 | 0.122 | 0.079 | 0.131 |